\documentclass[apjl]{emulateapj}
\usepackage[citebordercolor={0 .5 .5}]{hyperref}
\usepackage{multirow}
\usepackage{color}

\newcommand{\beq}{\begin{equation}}
\newcommand{\eeq}{\end{equation}}
\newcommand{\be}{\begin{equation}}
\newcommand{\ee}{\end{equation}}
\newcommand{\bea}{\begin{eqnarray}}
\newcommand{\eea}{\end{eqnarray}}
\newcommand{\bdi}{\begin{displaymath}}
\newcommand{\edi}{\end{displaymath}}

\newcommand{\rmicron}{$\,\micro$m}
\newcommand{\simstack}{{\sc simstack}}

\newcommand{\rmnum}[1]{\romannumeral #1}

\def\lsim{\,\lower2truept\hbox{${<\atop\hbox{\raise4truept\hbox{$\sim$}}}$}\,}
\def\gsim{\,\lower2truept\hbox{${>\atop\hbox{\raise4truept\hbox{$\sim$}}}$}\,}

\usepackage{amssymb,amsmath}
\usepackage{natbib}  
\usepackage[mediumspace,Gray,squaren]{SIunits}  
\usepackage{rotating}
\shorttitle{A New Accounting of the Cosmic Infrared Background}

\shortauthors{Viero, M.\,P. et al.}
\begin{document}
\title{HerMES: Current Cosmic Infrared Background Estimates Can be Explained by Known Galaxies and their Faint Companions at $z < 4$$^{\dagger}$}
\author{M.P.~Viero\altaffilmark{1,2},
L.~Moncelsi\altaffilmark{2},
R.F.~Quadri\altaffilmark{3},
M.~B{\'e}thermin\altaffilmark{4,5},
J.~Bock\altaffilmark{2,6},
D.~Burgarella\altaffilmark{7},
S.C.~Chapman\altaffilmark{8},
D.L.~Clements\altaffilmark{9},
A.~Conley\altaffilmark{10},
L.~Conversi\altaffilmark{11},
S.~Duivenvoorden\altaffilmark{12},
J.S.~Dunlop\altaffilmark{13},
D.~Farrah\altaffilmark{14},
A.~Franceschini\altaffilmark{15},
M.~Halpern\altaffilmark{16},
R.J.~Ivison\altaffilmark{17,13},
G.~Lagache\altaffilmark{7},
G.~Magdis\altaffilmark{18},
L.~Marchetti\altaffilmark{15},
J.~\'Alvarez-M\'arquez\altaffilmark{7},
G.~Marsden\altaffilmark{16},
S.J.~Oliver\altaffilmark{12},
M.J.~Page\altaffilmark{19},
I.~P{\'e}rez-Fournon\altaffilmark{20,21},
B.~Schulz\altaffilmark{2,22},
Douglas~Scott\altaffilmark{16},
I.~Valtchanov\altaffilmark{11},
J.D.~Vieira\altaffilmark{23,24},
L.~Wang\altaffilmark{25,26},
J.~Wardlow\altaffilmark{27},
M.~Zemcov\altaffilmark{2,6}}
\altaffiltext{1}{Kavli Institute for Particle Astrophysics and Cosmology, Stanford University, 382 Via Pueblo Mall, Stanford, CA 94305; ~ {email: marco.viero@stanford.edu}}
\altaffiltext{2}{California Institute of Technology, 1200 E. California Blvd., Pasadena, CA 91125}
\altaffiltext{3}{George P. and Cynthia Woods Mitchell Institute for Fundamental Physics and Astronomy, Department of Physics and Astronomy, Texas A\&M University, College Station, TX 77843}
\altaffiltext{4}{Laboratoire AIM-Paris-Saclay, CEA/DSM/Irfu - CNRS - Universit\'e Paris Diderot, CE-Saclay, pt courrier 131, F-91191 Gif-sur-Yvette, France}
\altaffiltext{5}{Institut d'Astrophysique Spatiale (IAS), b\^atiment 121, Universit\'e Paris-Sud 11 and CNRS (UMR 8617), 91405 Orsay, France}
\altaffiltext{6}{Jet Propulsion Laboratory, 4800 Oak Grove Drive, Pasadena, CA 91109}
\altaffiltext{7}{Laboratoire d'Astrophysique de Marseille - LAM, Universit\'e d'Aix-Marseille \& CNRS, UMR7326, 38 rue F. Joliot-Curie, 13388 Marseille Cedex 13, France}
\altaffiltext{8}{Dalhousie University, Department of Physics and Atmospheric Science, Coburg Road, Halifax, NS B3H 1A6, Canada}
\altaffiltext{9}{Astrophysics Group, Imperial College London, Blackett Laboratory, Prince Consort Road, London SW7 2AZ, UK}
\altaffiltext{10}{Center for Astrophysics and Space Astronomy 389-UCB, University of Colorado, Boulder, CO 80309}
\altaffiltext{11}{Herschel Science Centre, European Space Astronomy Centre, Villanueva de la Ca\~nada, 28691 Madrid, Spain}
\altaffiltext{12}{Astronomy Centre, Dept. of Physics \& Astronomy, University of Sussex, Brighton BN1 9QH, UK}
\altaffiltext{13}{Institute for Astronomy, University of Edinburgh, Royal Observatory, Blackford Hill, Edinburgh EH9 3HJ, UK}
\altaffiltext{14}{Department of Physics, Virginia Tech, Blacksburg, VA 24061}
\altaffiltext{15}{Dipartimento di Astronomia, Universit\`{a} di Padova, vicolo Osservatorio, 3, 35122 Padova, Italy}
\altaffiltext{16}{UK Astronomy Technology Centre, Royal Observatory, Blackford Hill, Edinburgh EH9 3HJ, UK}
\altaffiltext{17}{Department of Astrophysics, Denys Wilkinson Building, University of Oxford, Keble Road, Oxford OX1 3RH, UK}
\altaffiltext{18}{Department of Physics \& Astronomy, University of British Columbia, 6224 Agricultural Road, Vancouver, BC V6T~1Z1, Canada}
\altaffiltext{19}{Mullard Space Science Laboratory, University College London, Holmbury St. Mary, Dorking, Surrey RH5 6NT, UK}
\altaffiltext{20}{Instituto de Astrof{\'\i}sica de Canarias (IAC), E-38200 La Laguna, Tenerife, Spain}
\altaffiltext{21}{Departamento de Astrof{\'\i}sica, Universidad de La Laguna (ULL), E-38205 La Laguna, Tenerife, Spain}
\altaffiltext{22}{Infrared Processing and Analysis Center, MS 100-22, California Institute of Technology, JPL, Pasadena, CA 91125}
\altaffiltext{23}{Department of Physics, University of Illinois Urbana-Champaign, 1110 W. Green Street, Urbana, IL 61801}
\altaffiltext{24}{Astronomy Department, University of Illinois at Urbana-Champaign, 1002 W. Green Street, Urbana, IL 61801}
\altaffiltext{25}{SRON Netherlands Institute for Space Research, Landleven 12, 9747 AD, Groningen, Netherlands}
\altaffiltext{26}{Institute for Computational Cosmology, Department of Physics, University of Durham, South Road, Durham, DH1 3LE, UK}
\altaffiltext{27}{Dark Cosmology Centre, Niels Bohr Institute, University of Copenhagen, Juliane Maries Vej 30, 2100 Copenhagen, Denmark}
\begin{abstract}
We report contributions to cosmic infrared background (CIB) intensities originating from known galaxies and their faint companions at submillimeter wavelengths.   Using the publicly-available UltraVISTA catalog, and maps at 250, 350, and 500\rmicron\ from the \emph{Herschel} Multi-tiered Extragalactic Survey (HerMES),   
we perform a novel measurement that exploits the fact that uncatalogued sources may bias stacked flux densities --- particularly if the resolution of the image is poor --- and
intentionally smooth the images before stacking and summing intensities.    
By smoothing the maps we are capturing the contribution of faint (undetected in $K_S\sim 23.4$) sources that are physically associated, or \emph{correlated}, with the detected sources.  
We find that the cumulative CIB increases with increased smoothing, reaching $9.82\pm 0.78$, $5.77\pm 0.43$, and $2.32\pm 0.19\, \rm nW m^{-2} sr^{-1}$ at 250, 350, and 500\rmicron\ at $ 300\, \rm arcsec$ full width at half-maximum.  This corresponds to a fraction of the fiducial CIB of $0.94\pm 0.23$, $1.07\pm 0.31$, and $0.97\pm 0.26$ at 250, 350, and 500\rmicron,   {where the uncertainties are dominated by those of the absolute CIB}.  
We then propose, with a simple model combining parametric descriptions for stacked flux densities and stellar mass functions, that emission from galaxies with log($M/ \rm M_{\odot}) > 8.5$  can account for the most of the measured total intensities,  and argue against contributions from extended, diffuse emission.   
Finally, we discuss prospects for future survey instruments to improve the estimates of the absolute CIB levels, and observe any potentially remaining emission at $z > 4$. 
\end{abstract}


\section{Introduction}
\label{sec:intro}
\setcounter{footnote}{0}
Of all the light that has been emitted by stars,  about half has been absorbed by interstellar dust and thermally re-radiated at far-infrared to submillimeter wavelengths, appearing as a diffuse, extragalactic, cosmic infrared background spanning 1--1000\rmicron\ \citep[CIB;][]{hauser2001, dole2006}.  
Statistically characterizing the sources responsible for this background is necessary to gain a full understanding of galaxy formation and cosmology, and thus remains an ongoing pursuit.  

The CIB was first detected in spectroscopy with the Far Infrared Absolute Spectrophotometer \citep[FIRAS;][]{puget1996, mather1999}.  Observations of local starburst galaxies with the \emph{Infrared Astronomical Satellite} \citep[\emph{IRAS};][]{soifer1984}  showed that galaxies could emit a surprisingly large part of their energy in the far-infrared, and ground-based measurements later confirmed the existence of a rare population of extremely luminous submillimeter galaxies \citep[or SMGs;][]{blain2002}.  While these bright objects generated tremendous excitement,   and in fact constitute a significant fraction of the total star-formation rate density at $z > 3$ \citep[e.g.,][]{lefloch2005, murphy2011}, their low abundance only accounts for a small fraction of the total CIB.    

The arrival of the \emph{Herschel Space Observatory} --- whose instruments, PACS \citep[70, 100, and 160\rmicron;][]{poglitsch2010} and SPIRE \citep[250, 350, and 500\rmicron;][]{griffin2010}, bracket the peak of the thermal spectrum of dust emission --- brought the promise of directly detecting less luminous and far more numerous dusty star-forming galaxies (DSFGs).    
However,  source confusion resulting from the relatively large point-spread functions \citep[e.g.,][]{nguyen2010} limited the number of galaxies that could be individually resolved by PACS at 100 and 160\rmicron\ to 75\% \citep[][]{magnelli2013} and 74\% \citep[][]{berta2011} of the CIB, respectively, and  
by SPIRE at 250\rmicron\  to 15\%  \citep{bethermin2012b,oliver2012}.  
Statistical methods including stacking \citep[e.g.,][]{bethermin2012b,viero2013b} and \emph{P(D)} \citep[][]{glenn2010,berta2011}  performed  better, resolving  
70\% at 250, 350, and 500\rmicron\ for the former, and 
89\% and 70\% of the CIB at 100 and 250\rmicron\ for the latter, respectively.   

The origin of the rest remained unclear.  \citet[][]{viero2013b} suggested that the missing flux could be tied up in faint sources  --- faint either because they are low mass, at high redshift, or extremely dusty.  Another possible source is diffuse emission from the dust that is 
known to be distributed in the halos of galaxies \citep{menard2010,hildebrandt2013}, and could be heated by stripped stars \citep[e.g.,][]{tal2011, zemcov2014}.

Here we present a technique to demonstrate that most, if not all, of the CIB can be accounted for by the combined emission from galaxies detected in current near-infrared surveys, 
and their faint companion objects, at $z < 4$.  We show with a simple model that galaxies alone are the most plausible source of this signal,  and argue that any remaining CIB likely originates from galaxies at still higher redshifts.  

\section{Data}
\label{sec:data}
\subsection{UltraVISTA Catalog}
We perform our analysis on catalogs and images located in the {\sc COSMOS} field \citep[][]{scoville2007},   
centered at $\rm 10^{h}00^m26^s, +2^{\circ}13\arcmin 00\arcsec$.
We use the $K_{\rm S} = 23.4\, \rm (AB)$-selected, publicly-available\footnote{\url{http://www.strw.leidenuniv.nl/galaxyevolution/ULTRAVISTA}} 
catalog from \citet[][UltraVISTA]{muzzin2013b}, which consists of a $1.62\, \rm \deg^2$ subset of the full {\sc COSMOS} field, where both near-infrared and optical wavelengths are available (30 bands in all).    
The catalog contains photometric redshifts computed with EAZY \citep[][]{brammer2008}, and stellar masses computed with FAST \citep[][]{kriek2009a}.  
Galaxies are split into star-forming or quiescent based on their positions in the rest-frame $U-V$ vs.\@ $V-J$ color-color diagram 
 \citep[\emph{UVJ};][]{williams2009}. 
\subsection{Herschel/HerMES Submillimeter Maps}
We use submillimeter maps observed with the Spectral and Photometric Imaging REceiver \citep[SPIRE;][]{griffin2010} at 250, 350, and 500\rmicron\ from the Herschel Multi-tiered Extragalactic Survey \citep[HerMES;][]{oliver2012}. {\sc COSMOS} is a level 5 field, consisting of 4 repeat observations, to an instrumental depth of 15.9, 13.3, and 19.1\,mJy (5$\sigma$), with confusion adding an additional noise term of  24.0, 27.5, and 30.5\,mJy (5$\sigma$) at 250, 350, and 500\rmicron, respectively \citep{nguyen2010}.  
Absolute calibration is detailed in \citet{griffin2013}, with calibration uncertainties of $5\%$.   
Maps are made with 4\,arcsec pixels at all wavelengths 
using the {\sc SMAP} \citep{levenson2010, viero2013a} pipeline.  

SPIRE maps are chosen specifically for this study because its wavelengths probe the rest-frame peak of thermal dust emission at $z=1$--3 \citep{madau2014}, and because large-scale features can be reconstructed with minimal filtering \citep[][]{pascale2011}. 

\begin{figure*}[!t]
\centering
\includegraphics[width=1.0\textwidth]{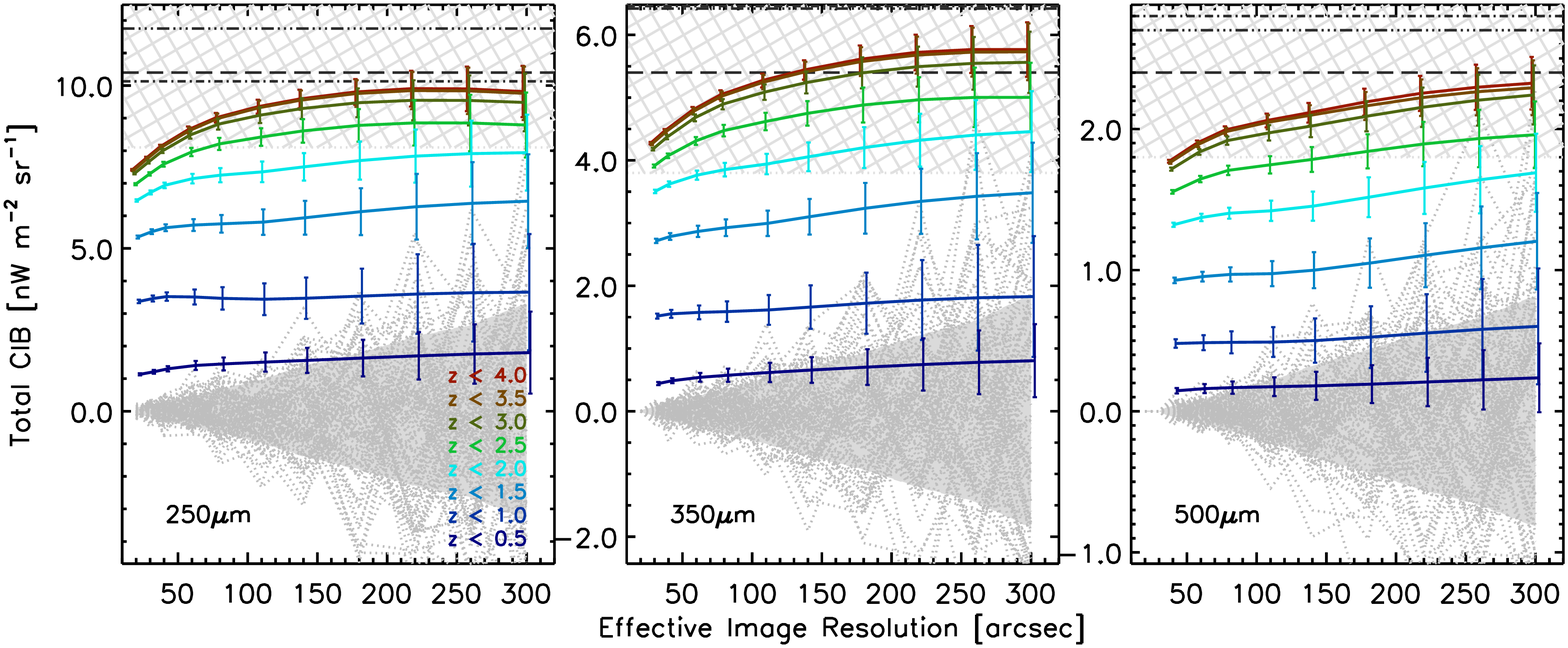}
\caption{
Cumulative measured CIB vs.\@ the size of the effective image resolution (in arcsec, FWHM) at 250 (left panel), 350 (center panel), and 500\rmicron\ (right panel).  
The \citet[][]{fixsen1998} FIRAS values and 1\,$\sigma$ errors are shown as dashed lines with gray hatched regions,  
while the \citet{lagache1999} FIRAS values are shown as 3-dot-dashed lines, and the  
\citet{bethermin2012b} model estimates  are shown as dot-dashed lines.  
Colors represent the sum over all bins up to the given $z$. 
Grey dotted lines show the full set of null tests, and shaded regions the 1\,$\sigma$ uncertainties, for the $z < 4$ case.   
The cumulative CIB vs.\@ effective resolution increases more rapidly at higher redshift, where the catalog in increasingly incomplete.  
The flattening of the curve at the highest redshift suggests that any potential remaining intensity lies at higher redshifts.  
}
\vspace{1mm}
\label{fig:tcib}
\end{figure*}

\section{Method}
\label{sec:method}
We now present a novel method to estimate the extent to which known sources and their faint companions contribute to the CIB. 
We do this by exploiting an inherent weakness of stacking: 
that stacking on images with poor angular resolution 
can result in a boosted (or biased) average flux density arising from faint, uncatalogued,  companion galaxies \citep[e.g.,][]{serjeant2008, fernandez-conde2010,kurczynski2010,heinis2013,viero2013b}.  
The trick is in recognizing that only \emph{correlated} (i.e.,  clustered) sources will bias the stacked flux density \citep[for an in-depth discussion see][]{marsden2009,viero2013b}, so that summed intensities estimated with increasingly smoothed maps places limits on the full intensity in a given redshift range.     
Meanwhile, emission that is not correlated --- say, emission coming from sources at redshifts greater than those of our catalog objects --- will not influence the primary measurement, except as a potential noise term.  
 
For this analysis we use the publicly-available \simstack\ code\footnote{\url{http://www.stanford.edu/~viero/downloads.html}}, which is described in detail in \citet{viero2013b}.  
Briefly, synthetic images of the sky are constructed from \emph{correlated} subsets of catalogs (i.e., objects in the same redshift range) with the assumption that galaxies that are physically similar --- in this case quiescent or star-forming galaxies within a stellar mass bin --- have comparable infrared luminosities and submillimeter flux densities.   Synthetic images are then convolved with the PSF of the instrument, and are fit simultaneously to the actual sky map to retrieve the mean flux densities of the subsample.  

To induce a bias we smooth the maps  by convolving them with Gaussians whose widths are the geometric differences of the nominal SPIRE and effective beams, i.e., $\sigma_{\rm smooth} =\sqrt{\sigma_{\rm eff}^2-\sigma_{\rm SPIRE}^2}$, where the nominal SPIRE resolutions are 17.5, 23.7, and 34.6\,arcsec full width at half-maximum (FWHM)  at 250, 350, and 500\rmicron, respectively, and the effective resolutions are 20, 30, 40, 60, 80, 110, 140, 180, 220, 260, and 300\,arcsec FWHM.  
Above 300\,arcsec, statistical uncertainties become consistent with zero (see Section~\ref{sec:results}).  
All synthetic images and sky maps are mean-subtracted before stacking.   

Following \citet{viero2013b} we split the sample into $8\times 8$ bins of redshift ($z=0$ to 4 with $\Delta z =0.5$) and stellar mass (5 star-forming and 3 quiescent), and stack the subsamples with \simstack\ at each smoothing scale.  Stacked flux densities are then color-corrected to account for the observed spectral shape of the sources  through the passbands, with temperatures taken from \citet[][Equation~18]{viero2013b}. 
Color corrections range between, 1.006--0.984, 0.993--0.977, and 1.007--0.991 at 250, 350, and 500\rmicron, respectively.    
 
\begin{figure*}[!t]
\centering
\hspace{-8mm}
\includegraphics[width=1.0\textwidth]{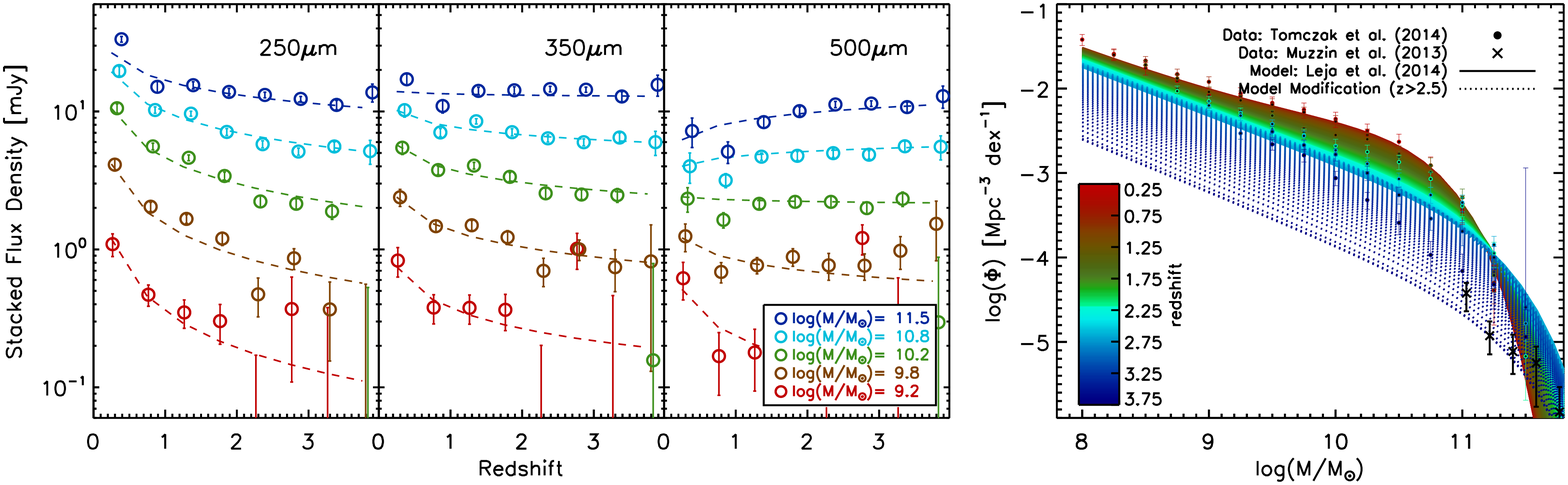}
\caption{
Left three panels: Stacked flux densities (measured with nominal PSFs) in divisions of stellar mass vs.\@ redshift for star-forming galaxies.  
Power-law model fits (described in Section~\ref{sec:model}) are shown as dotted lines. 
Right panel: Parametric model for the stellar mass functions of star-forming galaxies, which combines the parameterization of the \citet[][circles]{tomczak2014} data by \citet[][solid lines]{leja2015} at $z \leq 2.5$, and at modification at $z>2.5$ (dashed lines) which interpolates to measurements from \citet[][exes]{muzzin2013}. 
}
\vspace{1mm}
\label{fig:stackedflux}
\end{figure*}

We use simulations to check that stacking and smoothing with a Gaussian, as opposed to the measured SPIRE beam or any other kernel, is a reasonable approximation, finding a bias of less than 4\% for the largest smoothing kernel, which we include in the reported errors.  
We test that the method does not introduce unintended biases by performing 100 null tests for each set of synthetic images and maps.  
Null tests involve running the identical stacking pipeline with the same binning of sources, but after randomizing the positions of the sources in the catalogs. 
Because the map and images are mean-subtracted, we expect the stacked flux densities of the null tests to be consistent with zero. In Section~\ref{sec:results} we show that this is the case.
    
We note that this method has the advantage that missing sources are not double-counted,  
meaning that the flux density from a single missing object will be distributed among the synthetic images, 
rather than appearing multiple times (as would be the case in thumbnail stacking).  
Also note that stacked flux densities are intentionally \emph{not} corrected for completeness because it is precisely the flux densities of the missing (i.e., incomplete, but similarly applies to misclassified AGN or DSFGs) sources that we are attempting to measure by degrading the effective resolution of the map.   
As a result, this technique is not limited to CIB studies, but is applicable to any study where estimates of the level of faint, correlated emission are in question. 

\section{Results}
\label{sec:results}
We report total intensities of our stacking measurement  as the cumulative sum  over redshift (from $z = 0$ to the redshifts  labeled with different color lines) vs.\@ the effective resolution of the image, in Figure~\ref{fig:tcib}\footnote{Tabulated values for the intensities in Figure~1 can be found at \url{https://web.stanford.edu/\~viero/downloads.html}}.  
At lower redshifts ($z<1$) the fractional CIB that is measured increases weakly with increasing effective beam size, which is expected given that the completeness of the catalog at these redshifts is near unity for stellar masses log($M/ \rm M_{\odot}) \ge 9$.  
Conversely, at higher redshifts the fractional CIB that is measured increases rapidly with increased smoothing.  
The maximum CIB we resolve is $9.82\pm 0.78$, $5.77\pm 0.43$, $2.32\pm 0.19\, \rm nW m^{-2} sr^{-1}$ at 250, 350, and 500\rmicron.  
We note slightly different behaviors between the three bands; particularly at 500\rmicron, which appears to have not converged, and may be indicative of higher redshift contributions to the CIB.  
 
Estimates of the exact measured fraction of the absolute CIB are limited by the uncertainties in the reported absolute values of the CIB  derived from FIRAS spectra by \citet[][]{fixsen1998} and \citet[][]{lagache1999}, which range between 22\% and 30\%.  The \citet[][]{fixsen1998}  levels  ($10.4\pm 2.3$, $5.4\pm 1.6$, $2.6\pm 0.6$; hereafter chosen to represent the fiducial values) are shown as dashed lines in Figure~\ref{fig:tcib}, and the \citet[][]{lagache1999}  levels ($11.8\pm 2.9$, $6.4\pm 1.6$, $2.7\pm 0.7$)   are shown as 3-dot-dashed lines.  
Additionally, \citet[][dot-dashed lines]{bethermin2012b} provide estimates of the total CIB by extrapolating their measured counts, finding they agreed with FIRAS with similarly large uncertainties.  
In total we find that the fraction of the  \citet{fixsen1998} fiducial CIB  we resolve is $0.94\pm 0.23$, $1.07\pm 0.31$, and $0.97\pm 0.26$ at 250, 350, and 500\rmicron, respectively.  

The full set of 100 null tests at each effective beam size are shown as gray dotted lines in Figure~\ref{fig:tcib}, with the 1\,$\sigma$ limits represented by the shaded grey region.  For clarity only the tests for $z < 4$ are shown.   
As expected, the results of the null tests fluctuate around zero, with the magnitude of that fluctuation increasing with increasing effective beam size.  
Additional systematic uncertainties include calibration and beam area uncertainties, and cosmic variance, which is estimated following \citet{moster2011}. 

\begin{figure}[!t]
\centering
\vspace{-1mm}
\hspace{-7.0mm}
\includegraphics[width=0.51\textwidth]{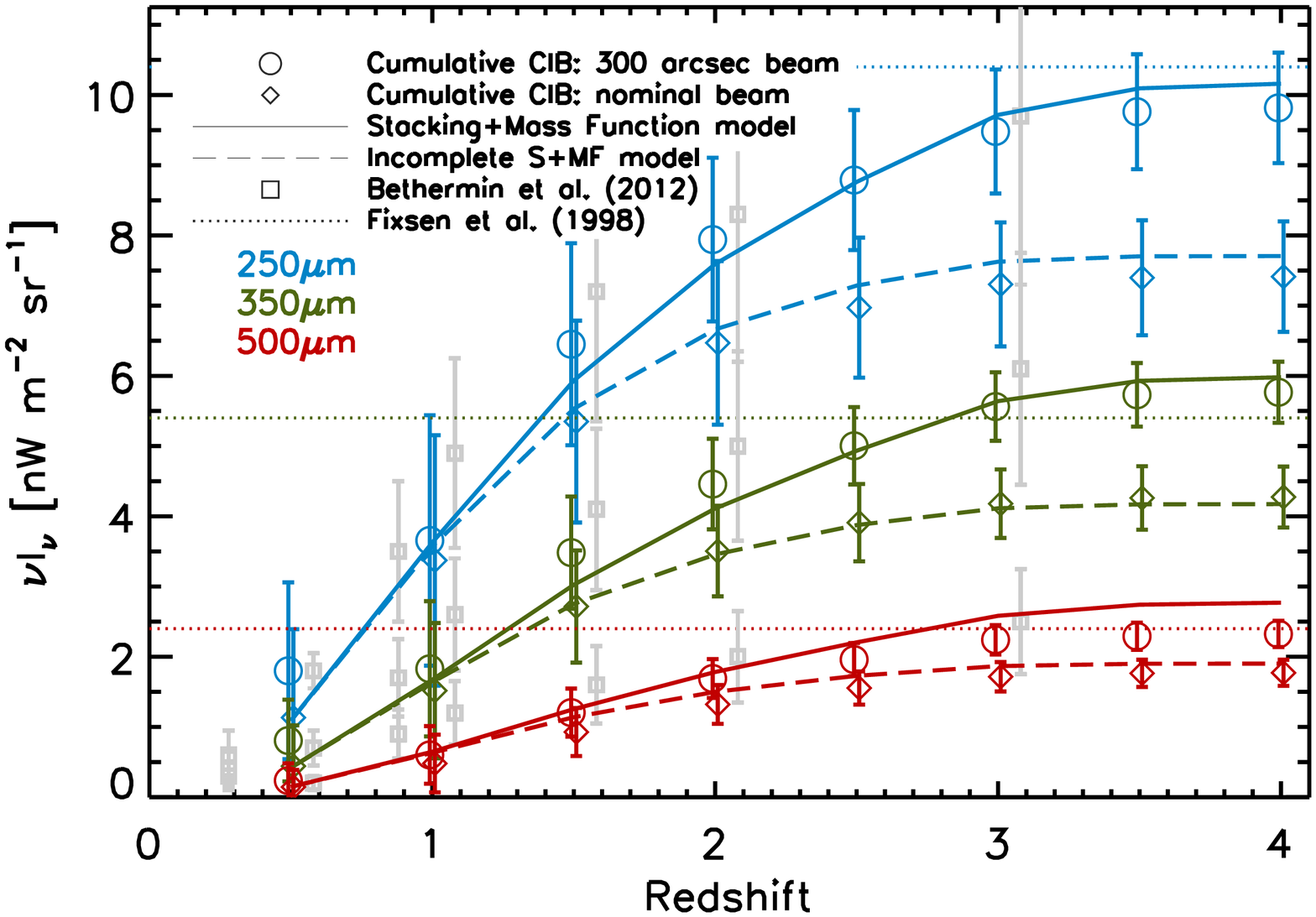}
\caption{
Measurements (open circles) and models (solid lines) for cumulative CIB intensities vs.\@ redshift.  
Measurements are the cumulative sums for images stacked at the highest effective beam size of 300\,arcsec FWHM.  
The models, described in Section~\ref{sec:model}, combine parametric descriptions for the stacked flux densities (at the native resolution of the images) and the stellar mass functions (see Figure~\ref{fig:stackedflux}).  
Extrapolated counts from \citet{bethermin2012b} are shown as gray squares.  
Also shown are cumulative CIB intensities vs.\@ redshift on stacks made with the native (i.e., non-smoothed) SPIRE images (diamonds), and the models {after they have been modified to reflect the incompleteness of the actual catalog} (dashed lines).  
}
\vspace{1.5mm}
\label{fig:didz}
\end{figure}

\section{Discussion}
\label{sec:discussion}
The CIB can be divided into three components: (\rmnum{1}) the contribution from the sources in the parent $K_S \lsim 23.4$ catalog; (\rmnum{2}) the contribution from correlated companion sources that for some reason do not make the catalog cut but are recovered with our smoothing-stacking method; and (\rmnum{3}) uncorrelated emission, including and likely dominated by sources lying at $z  > 4$.   

It is interesting to consider the nature of (\rmnum{2}), the undetected sources that are captured by our smoothing procedure:  are they very low-mass galaxies; massive but dusty galaxies that evade detection in current deep $K$-selected catalogs; or some other unknown component? 

We now propose --- through a simple model that combines a parametric description of the stellar mass function with simple fits to nominal stacked flux densities --- that the galaxies that make up the stellar mass function are alone able to describe our measurement, and that by extension the missing CIB can reasonably be attributed to  galaxies in the low-mass end of the stellar mass function.

\subsection{A Model of the CIB}
\label{sec:model}
The first component of the model adopts the \citet[][Equation~1]{leja2015} parameterization of the \citet{tomczak2014} stellar mass function.  
The rightmost panel of Figure~\ref{fig:stackedflux} illustrates the performance of the \citet{leja2015}  model (solid lines) against the  \citet{tomczak2014} data (circles).
We find that it diverges from the measurements at higher redshifts and so we add a modification at $z>2.5$ (dotted lines) by
simply interpolating between the model and the measured stellar mass functions of \citet{muzzin2013} to $z=4$.  
We check that the exact value of faint-end slope at high-$z$ negligibly affects the integral, and set it to -1.6.  

Similarly, the second component of the model consists of power-law fits to the stacked flux densities vs.\@ lookback time for each stellar mass bin, with the added condition that the mass dependence of the slopes and offsets themselves follow smooth functions. The three left panels of Figure~\ref{fig:stackedflux} show the stacked flux densities and best-fit power laws as open circles with error bars, and dotted lines, respectively.   
Finally, the model is integrated over the redshift range $z=0.1$--4, and stellar mass range log($M/ \rm M_{\odot}) = 8$--14; although the contribution from galaxies below log($M/ \rm M_{\odot}) < 8.5$ is found to be negligible.  

Figure~\ref{fig:didz} compares the resulting model (solid lines) with the full CIB measurements (open circles) at all three wavelengths.  Note that model is not a fit, and yet describes the measurement remarkably well,  demonstrating that the faint-end of the mass function can plausibly explain the recovered CIB.  
Also in good agreement are the \citet[][gray squares]{bethermin2012b} estimates derived through extrapolation of their number counts.  We note some tension with the \citet[][]{lagache2014} intensity at 350\rmicron\ ($7.7\pm 0.2\, \rm nW m^{-2}$), which is an output of their halo model fit to CIB power spectra; although their 550\rmicron\ estimate from the same model is in relative agreement ($2.3\pm 0.1\, \rm nW m^{-2}$).  

Also shown (as diamonds) are measurements made when stacking on the native SPIRE images (i.e., no smoothing).  
They are again compared to the model (dashed lines), but this time the model is modified so that the low-mass limit of the integral over the mass function effectively begins at higher masses with increasing redshift, reflecting the completeness behavior of the catalog.  
The difference between the solid and dashed lines can be interpreted as a measure of how much of the CIB originates from the parent sample, and how much is from companion sources not detected at this $K_S$ limit. 
If, for arguments sake, the catalog were 100\% complete at all stellar masses and all redshifts, then the dashed and solid lines would overlap.   

Arguments for additional, diffuse, sources of CIB; in particular dust in the extended halos of galaxies \citep{menard2010}, are thus disfavored. 

\subsection{A CIB beyond $z$ of 4?}
\label{sec:highz}
While our measurements are consistent with the fiducial levels of the total CIB, 
the existing uncertainties on the absolute level are so large that it is difficult to convincingly estimate how much CIB is still unresolved.  
Any missing CIB is more likely to occur at longer wavelengths, which are more sensitive to higher redshifts \citep[because of the negative $K$-correction;][]{blain2003}, from where we would expect uncorrelated emission to originate.  

Indeed, several exceptional ULIRG-like galaxies have been identified at $z > 4$, albeit with low abundances \citep[e.g.,][]{riechers2013,vieira2013, dowell2014,swinbank2014}.  However,  extrapolations of the contribution of ULIRGs to the star-formation rate density at $z > 4$ \citep[e.g.,][]{murphy2011,viero2013b} points to them dominating the far-infrared emission at this epoch.  
Determining the relative levels that they and less luminous galaxies contribute is a key question going forward. 

To resolve these high-$z$ questions, more data are required; particularly: \rmnum{1}) an update of the absolute CIB; \rmnum{2}) deeper catalogs with stellar masses and redshifts to redshifts greater than 4; \rmnum{3}) submillimeter surveys (350 to 1000\rmicron) with large angular-scale fidelity and smaller PSFs in the regions of those deep catalogs --- particularly at longer wavelengths which are more sensitive to galaxies at higher redshifts.  
The former can only be achieved with  space-based missions to measure the DC level above atmospheric foregrounds. 
The second requirement is steadily growing from multiple current or upcoming efforts (e.g., SDSS/BOSS, CANDLES, DES, LSST), although the photometric redshifts for very dusty galaxies may remain quite uncertain \citep[see][]{spitler2014}.  
The last point will require a ground-based submillimeter observatory similar in scope to CCAT \citep[][]{sebring2006}.    
\section{Conclusion}
\label{sec:conclusion}
We find that most of the CIB at  250, 350, and 500\rmicron\  can be accounted for by galaxies detected in current near-infrared surveys of moderate depth ($K_{\rm S}\approx 23.4$), and galaxies correlated with them.  
We report total intensities of $9.82\pm 0.78$, $5.77\pm 0.43$, $2.32\pm 0.19\, \rm nW m^{-2} sr^{-1}$ at 250, 350, and 500\rmicron, which corresponds to a fraction of the \citet{fixsen1998} absolute CIB of $0.94\pm 0.23$, $1.07\pm 0.31$, and $0.97\pm 0.26$. 

We find that a simple model combining parametric descriptions for the stellar mass function and stacked flux densities for log($M/ \rm M_{\odot}) > 8.5$ and $z < 4$ is able to convincingly reproduce the measurements, 
which supports the argument that the sources in the faint-end of the mass function make up the previously missing CIB.  We note that emission from objects that are not in the catalog, \emph{and} is either uncorrelated with catalog objects or exists at scales greater than 300\,arcsec, would be missed.  However unlikely, this cannot be ruled out without a better absolute measurement to compare against.  

Finally, we propose that any remaining CIB likely originates from galaxies at $z > 4$, and if so should be detectable at submillimeter and millimeter wavelengths.  

\begin{acknowledgments}
\section*{Acknowledgments}
MPV warmly thanks Charlotte Clarke, Pete Hurley, Seb Oliver, and the University of Sussex for their hospitality during the development of this study; and Phil Hopkins for valuable discussions of the $z>4$ Universe.  
We also thank the anonymous referee, whose careful comments greatly improved this paper.  
SPIRE has been developed by a consortium of institutes led
by Cardiff Univ. (UK) and including: Univ. Lethbridge (Canada);
NAOC (China); CEA, LAM (France); IFSI, Univ. Padua (Italy);
IAC (Spain); Stockholm Observatory (Sweden); Imperial College
London, RAL, UCL-MSSL, UKATC, Univ. Sussex (UK); and Caltech,
JPL, NHSC, Univ. Colorado (USA). This development has been
supported by national funding agencies: CSA (Canada); NAOC
(China); CEA, CNES, CNRS (France); ASI (Italy); MCINN (Spain);
SNSB (Sweden); STFC, UKSA (UK); and NASA (USA). 
The SPIRE data for this paper were obtained as a part of proposal {\tt KPGT\_soliver\_1}, with images made using the following OBSIDs: 1342222819-26, 1342222846-54, 1342222879-80, 1342222897-901.
\end{acknowledgments}



\end{document}